\documentclass[useAMS,usenatbib]{mn2e}

\usepackage{graphicx}

\title[Abundances in starbursts from C\,IV\,$\lambda 1549$]{On the 
reliability of C\,{\sc iv}\,$\lambda 1549$ as an 
abundance indicator for high redshift star-forming galaxies}
\author[Crowther, Prinja, Pettini and Steidel]{
Paul A. Crowther$^{1}$, Raman K. Prinja$^{2}$, Max Pettini$^{3}$ 
and Charles C. Steidel$^{4}$\\
$^{1}$Dept of Physics \& Astronomy, University of Sheffield, Hounsfield 
Road, Sheffield, S3 7RH, UK \\
$^{2}$Dept of Physics \& Astronomy, University College London, Gower 
Street, London, WC1E 6BT, UK \\
$^{3}$ Institute of Astronomy, Madingley Road, Cambridge, CB3 OHA, UK\\
$^{4}$ California Institute of Technology, MS 105-24, Pasadena, CA 91125, 
USA
}
\begin{document}

\date{Received: 2006 January 24; Accepted 2006 February 7}

\pagerange{\pageref{firstpage}--\pageref{lastpage}} \pubyear{2006}

\maketitle

\label{firstpage}

\begin{abstract}
We reconsider the use of the  equivalent width 
of C\,{\sc iv}\,$\lambda 1549$, EW(C\,{\sc iv}),
as an indicator of the oxygen abundance in star-forming galaxies, 
as proposed by Heckman et al. for nearby starbursts. 
We refine the local calibration of EW(C\,{\sc iv}) vs. $\log {\rm (O/H)}$
by using a restricted wavelength window which minimises blending with 
interstellar absorption lines. When applied to the \emph{stellar
component only} of the complex C\,{\sc iv}\,$\lambda 1549$ 
features in two high redshift galaxies with good quality spectra,
MS\,1512$-$cB58 ($z = 2.7268$) and Q1307-BM1163 ($z = 1.4105$),
the local calibration gives values of the oxygen abundance which are
in good agreement with other metallicity determinations based on
nebular emission and interstellar absorption lines.
Our main conclusion is that for this method to give reliable results
at high redshifts, it should only be used on data of sufficiently 
high spectral resolution ($R \ga 1000$) for stellar and interstellar 
C\,{\sc iv} components to be clearly separated. 
Oxygen abundances will be systematically overestimated if the 
local calibration is applied to spectra of high $z$ galaxies 
obtained with the low resolving powers ($R \simeq 200-300$)
of many current wide field surveys. 
It will also be necessary to understand better the causes
of the scatter in the local relation, before we can be confident 
of inferences from it at high $z$.
\end{abstract}


\section{Introduction}

Great progress has been made recently towards 
establishing both the star formation history 
(e.g. Madau et al. 1996; Hopkins 2004; 
Bunker et al. 2004 and references therein), 
and the chemical enrichment history 
(e.g. Kewley \& Kobulnicky 2005; Pettini 2006 
and references therein)  of the universe. 
The latter, whilst observationally more challenging, 
represents a powerful means of assessing how metallicity 
responds to star formation from $z \sim$ 5 through 
to the present day.

As discussed by Pettini (2006), various metallicity diagnostics
have been applied to the analysis of the spectra of high redshift
galaxies. Those most widely used so far are based on 
rest-frame optical emission lines from H\,{\sc ii} regions
(e.g. Pagel et al. 1979; Kewley \& Dopita 2002; 
Pettini \& Pagel 2004).
In a few exceptionally bright, or gravitationally lensed, 
galaxies these nebular metallicity measures have
been supplemented by estimates derived from the
strengths of photospheric lines from OB stars at rest-frame
ultraviolet (UV) wavelengths
(e.g. Leitherer et al. 2001; Rix et al. 2004).

These methods have their limitations, however.
The accurate measurement of shallow, stellar photospheric
features requires spectra of higher signal-to-noise ratio
(S/N) than generally attainable at present 
(see, for example, Erb et al. 2006). 
The nebular emission lines, on the other hand, are only
accessible (from the ground) at redshifts which place them 
within gaps between the numerous OH emission lines from the 
night sky which mar the near-IR spectral windows.
The most severe limitation of the nebular abundance 
diagnostics is that they cease to be applicable at redshifts
$z \ga 3.4$, as [O\,{\sc iii}]\,$\lambda 5007$ is 
redshifted beyond the $K$-band window. Their application
to the increasingly large numbers of galaxies at
$3.4 <  z <  6.5$ (e.g.  Iwata et al. 2005)
will have to await the advent of the 
\emph{James Webb Space Telescope (JWST)} in the next decade.

With thousands of UV spectra of galaxies at $z \ga 1.5$ now
available (e.g. Steidel et al. 2003, 2004; Le F\`{e}vre et al. 2005), 
attention naturally turns to searching for useful
metallicity measures in the rest-frame UV spectral region. 
The work by Rix et al. (2004) identified the strong
P Cygni lines of C\,{\sc iv}\,$\lambda 1549$ and
Si\,{\sc iv}\,$\lambda 1397$\footnote{For convenience we use 
the multiplet wavelengths. The lines are actually blended doublets
at wavelengths $\lambda\lambda 1548.20,1550.78$ and
$\lambda\lambda 1393.76, 1402.77$ respectively.} 
as the most suitable for this purpose. These features
originate in the winds of the most luminous OB stars
and their strengths are reduced at subsolar
metallicities, reflecting the lower mass-loss rates and wind
terminal velocities of these stars at $Z < Z_\odot$
(Prinja \& Crowther 1998; Leitherer et al. 2001, Vink et al. 2001).
Storchi-Bergmann, Calzetti, \& Kinney (1994) were the first to confirm
observationally a relationship between the equivalent widths
of C\,{\sc iv}\,$\lambda 1549$ and
Si\,{\sc iv}\,$\lambda 1397$ in the integrated spectra
of local star-forming galaxies recorded with 
the \emph{International Ultraviolet Explorer (IUE)} satellite
and their oxygen abundance (determined from nebular
emission lines). The correlation was later confirmed 
and quantified by the analyses of Heckman et al. (1998) 
and Mehlert et al. (2002). 

Heckman et al. (1998) emphasised the fact that the \emph{IUE} 
spectra, taken through the large entrance aperture, sample 
comparable physical scales to those encompassed by 
ground-based spectroscopic observations of high redshift 
galaxies and should therefore provide a set of local templates 
suitable for the interpretation of distant galaxies. Thus, their
calibration of the equivalent widths of   
C\,{\sc iv}\,$\lambda 1549$ and
Si\,{\sc iv}\,$\lambda 1397$ vs. O/H has been used to
infer the metallicity of galaxies at $z > 1.5$ 
(e.g. Mehlert et al. 2002, 2005). 
Given that these wind lines are probably our \emph{only} 
means of estimating abundances in galaxies at $z \ga 3.5$,
as explained above, we thought it important to reassess
the local calibrations.   
Specifically, we examine how well the
calibrations perform in the cases of two high redshift
galaxies with exceptionally high S/N UV spectra and 
for which a number of independent metallicity 
determinations have been reported. 
We pay particular attention to the effect which spectral
resolution and blending with interstellar absorption
lines have on the determination of abundances from
UV wind lines.

%
%
\begin{figure}
\begin{center}
\includegraphics[width=0.9\columnwidth,clip,angle=0]{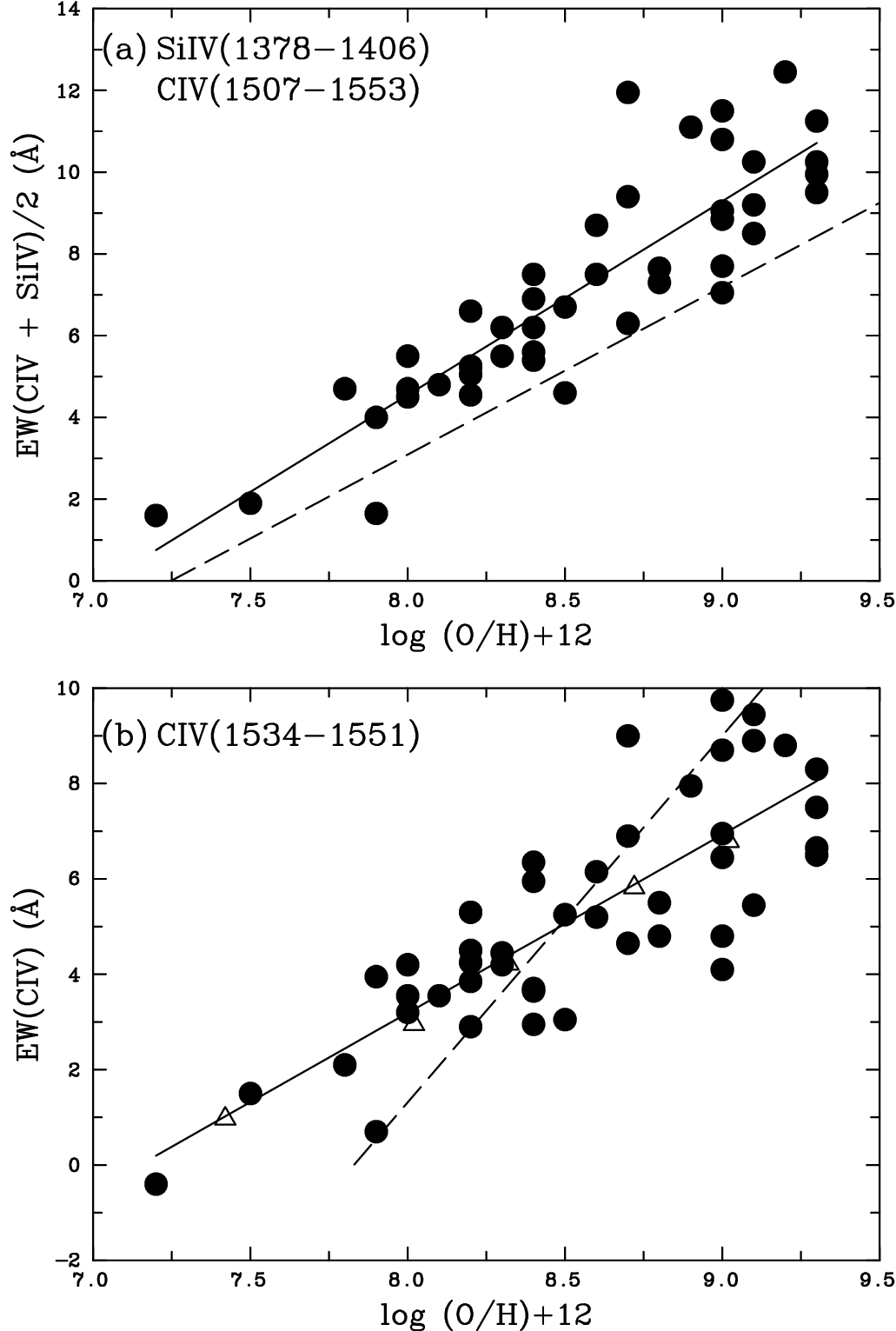}
\end{center}
\caption{(a) Average equivalent widths of the
C\,{\sc iv}\,$\lambda 1549$ and
Si\,{\sc iv}\,$\lambda 1397$ lines, 
measured from low resolution 
{\it IUE\/} spectra of local starburst galaxies,
vs. the oxygen abundance. Although we
followed the procedure outlined by 
Heckman et al. (1998) and the measurements refer
to the same sample of galaxies, we find a 
$1-2$\,\AA\ offset between our best fit
(continuous line) and theirs (dashed line);
(b) Equivalent width of C\,{\sc iv}\,$\lambda 1549$ 
versus oxygen abundance for the same sample
of galaxies, but now using a restricted spectral window,
$1534-1551$\,\AA, which excludes the 
Si\,{\sc ii}\,$\lambda\lambda 1527, 1533$
interstellar absorption lines. 
The continuous line is the best fit to our measurements,
given by Eqn.~(1), while the dashed line shows the relationship
derived by Mehlert et al. (2002) using a wider integration range
to measure EW(C\,{\sc iv}).
The open triangles show the values we measured from
the fully synthetic spectra of Rix et al. (2004).\label{mean}}
\end{figure}

\section{Local Starbursts}

We begin by attempting to reproduce the local calibration by
Heckman et al. (1998). To this end, we remeasured the
equivalent widths of the C\,{\sc iv}\,$\lambda 1549$ and
Si\,{\sc iv}\,$\lambda 1397$ doublets in the 
same subsample of 45 star-forming galaxies from the
Kinney et al. (1993) low resolution {\it IUE} galaxy atlas,
performing the equivalent width integrations over
the same spectral windows as those used by Heckman et al.,
1507--1553\,\AA\ and 1378--1406\,\AA\ for 
C\,{\sc iv} and Si\,{\sc iv} respectively.
The measurements were made independently by 
two of us (PAC and RKP) using the Starlink {\sc dipso} software
package (Howarth et al. 2003) and averaged to give
the quantity EW(C\,{\sc iv}  +  Si\,{\sc iv})/2 which
can then be directly compared with the same measurement
reported by Heckman et al., as in Fig.~\ref{mean}a.
For the purpose of this comparison we adopted the same
galaxy redshifts and oxygen abundances as Heckman et al.,
even though there are indications from more recent work
(Garnett et al. 2004a; Garnett, Kennicutt \& Bresolin 2004b; 
Bresolin, Garnett \& Kennicutt 2004; Bresolin et al. 2005)
than the R23 index of Pagel et al. (1979), on which the values
of (O/H) in Fig.~\ref{mean} are based, overestimates the oxygen
abundance at high (i.e. apparently super-solar) metallicities.

As can be seen from Fig.~\ref{mean}a, we do confirm the increase
of EW(C\,{\sc iv} + Si\,{\sc iv})/2 with $\log {\rm (O/H)} +12$
reported by Heckman et al., but also find a 1--2\,\AA\
offset between their best fit and ours. This is clearly a concern, 
further compounded by the relative large dispersion,
with standard deviation $\sigma \simeq 1$\,\AA, between
the values of EW(C\,{\sc iv} + Si\,{\sc iv})/2 measured
independently by two of the  authors.
The origin of these differences is unclear;
they may reflect different placements of the continuum level,
but in any case they are certainly a reason for
caution in the application of the Heckman et al.'s 
relation.

The  Si\,{\sc iv}\,$\lambda 1397$ wind line is nearly always
weaker than C\,{\sc iv}\,$\lambda 1549$ and its equivalent
width more difficult to reliably measure. We therefore chose to
concentrate on C\,{\sc iv}\,$\lambda 1549$ alone
(as done by Mehlert et al. 2002), and 
remeasured EW(C\,{\sc iv}), this time over a 
more restricted wavelength interval,  
$1534-1551$\,\AA, which corresponds to C\,{\sc iv} $\lambda$1548.20
wind  velocities from --2750 to +550\,km~s$^{-1}$ and avoids the 
nearby interstellar absorption lines 
Si\,{\sc ii}\,$\lambda\lambda 1526.71, 1533.43$.
Again the values of EW(C\,{\sc iv})
were measured independently by two of us
and averaged; in this case we found the scatter between the
two sets of measure to be considerably smaller than that
of the earlier measurements, with  $\sigma \simeq 0.2$\,\AA.
Values of EW(C\,{\sc iv}) are plotted in 
Fig.~\ref{mean}b vs. $\log {\rm O/H}$, together
with our line of best fit which satisfies the equation:
\begin{equation}\label{eqn1}
 \log {\rm (O/H)} + 12  = 7.15 + {\rm EW(C\,IV)}/3.74 
\end{equation}
where EW(C\,{\sc iv}) is in \AA.

\begin{figure}
\begin{center}
\includegraphics[width=0.9\columnwidth,clip,angle=0]{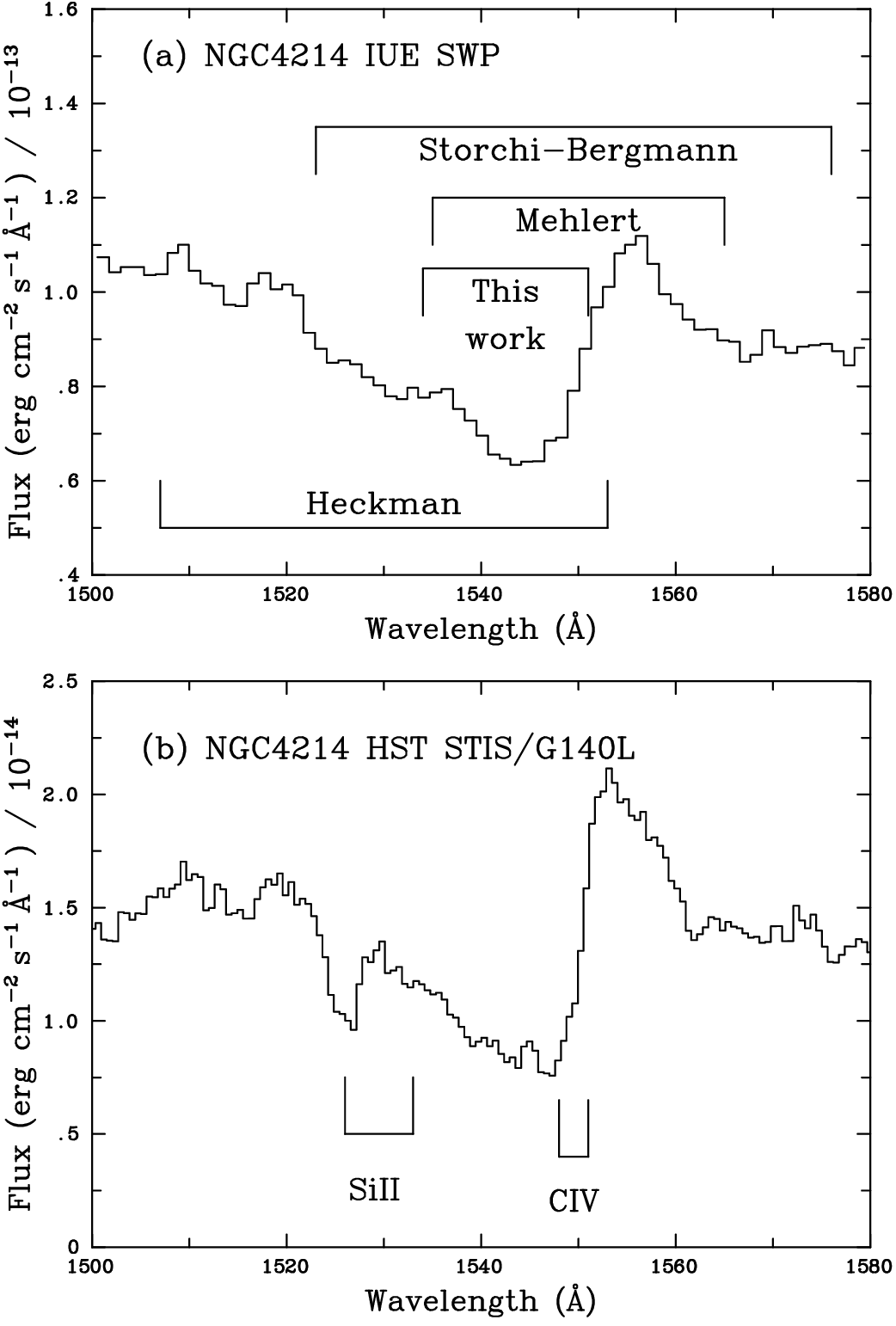}
\end{center}
\caption{(a) Portion of the {\it IUE}/SWP spectrum of the starburst galaxy 
NGC~4214 (corrected to the rest-frame of the galaxy)
from the Kinney et al. (1993) atlas, in the region of the
C\,{\sc iv}\,$\lambda 1549$ line. 
The resolving power is $R\simeq 250$.
Also shown are the windows used by different
authors to measure EW(C\,{\sc iv}).  (b) The same portion
of the higher resolution ($R \simeq 600$) archival
{\it HST}/STIS G140L spectrum of 
a bright UV knot within the galaxy 
(NGC~4214-1; Chandar, et al. 2004).
The locations of the narrow interstellar lines of Si\,{\sc ii} and 
C\,{\sc iv} are indicated.
\label{ngc4214}}
\end{figure}

Also shown with a dash line in Fig.~\ref{mean}b is the 
relationship between EW(C\,{\sc iv}) and $\log {\rm (O/H)} +12$
proposed by Mehlert et al. (2002; their Eqn.~3) which is significantly
steeper than that found here. Undoubtedly this is due, at
least in part, to the wider spectral window adopted by
these authors for their equivalent width measurements:
$1535-1565$\,\AA, corresponding to velocities of
$-2550$ to +3250\,km~s$^{-1}$. Thus, the Mehlert et al.
measures refer to the combined (emission plus absorption)
equivalent width of the P Cygni feature, as do those by 
Storchi-Bergmann et al. (1994). The windows chosen by
all the analyses of the C\,{\sc iv}\,$\lambda 1549$ line
mentioned above are compared in Fig.~\ref{ngc4214}a.

Returning to Fig.~\ref{mean}b, we see that, apart from the 
ambiguity due to different wavelength ranges over which
EW(C\,{\sc iv}) is measured, there appears to be considerable
scatter of the data about the line of best fit. This could be due 
to intrinsic dispersion, reflecting the evolutionary status
of the starbursts; to errors in both the measurement of 
EW(C\,{\sc iv})---the \emph{IUE} spectra are of limited S/N---and
the determination of the oxygen abundance; and to 
varying strength of interstellar C\,{\sc iv} absorption which
is buried within the stellar P Cygni profile at the low resolution
($R\simeq 250$) of the \emph{IUE} spectra. It is interesting, however, 
that our line of best fit in Fig.~\ref{mean}b is in excellent
agreement with the values (open triangles) we measure from 
the fully synthetic UV spectra produced by Rix et al. (2004) for 
the standard case of continuous star formation with a Salpeter 
Initial Mass Function (IMF). These are purely stellar synthetic spectra 
and so do not include any interstellar absorption.

%
%

The degree of contamination by interstellar absorption lines
becomes clearer when we compare the low resolution \emph{IUE} 
spectra with those obtained with the \emph{Hubble Space Telescope (HST)},
although the latter normally do not sample the whole galaxy
(and thus would generally \emph{under}estimate the interstellar contamination).
An example is reproduced in Fig.~\ref{ngc4214}a for the
Wolf-Rayet galaxy NGC~4214. From the 
low resolution \emph{IUE} spectrum, we measure 
EW(C\,{\sc iv})\,=\,5.2\,\AA\ within the spectral range suggested
by Heckman et al. (1998), and EW(C\,{\sc iv})\,=\,4.2\,\AA\
using our smaller integration window which does not include
the interstellar Si\,{\sc ii}\,$\lambda\lambda 1527, 1533$ absorption lines.
The higher resolution \emph{HST} STIS spectrum of a bright
knot within the galaxy 
(NGC~4214-1, Chandar, Leitherer, \& Tremonti 2004)
shows clearly Si\,{\sc ii}\,$\lambda 1527$ 
and what, in this instance, appears to be only
a weak contribution by interstellar C\,{\sc iv}\,$\lambda 1549$ 
to the P Cygni line.

\section{High-redshift galaxies}

\subsection{Abundances from medium resolution rest-frame UV spectroscopy}

In the case of most star-forming galaxies at high
redshifts,  however,  interstellar absorption 
{\emph is} a significant contributor
to the integrated equivalent width of C\,{\sc iv}\,$\lambda 1549$,
as can be readily appreciated from inspection of the composite
spectra of galaxies at $z = 2 - 3$ published by Shapley et al. (2003)
and Erb et al. (2006).

%
%
\begin{figure}
\begin{center}
\includegraphics[width=0.9\columnwidth,clip,angle=0]{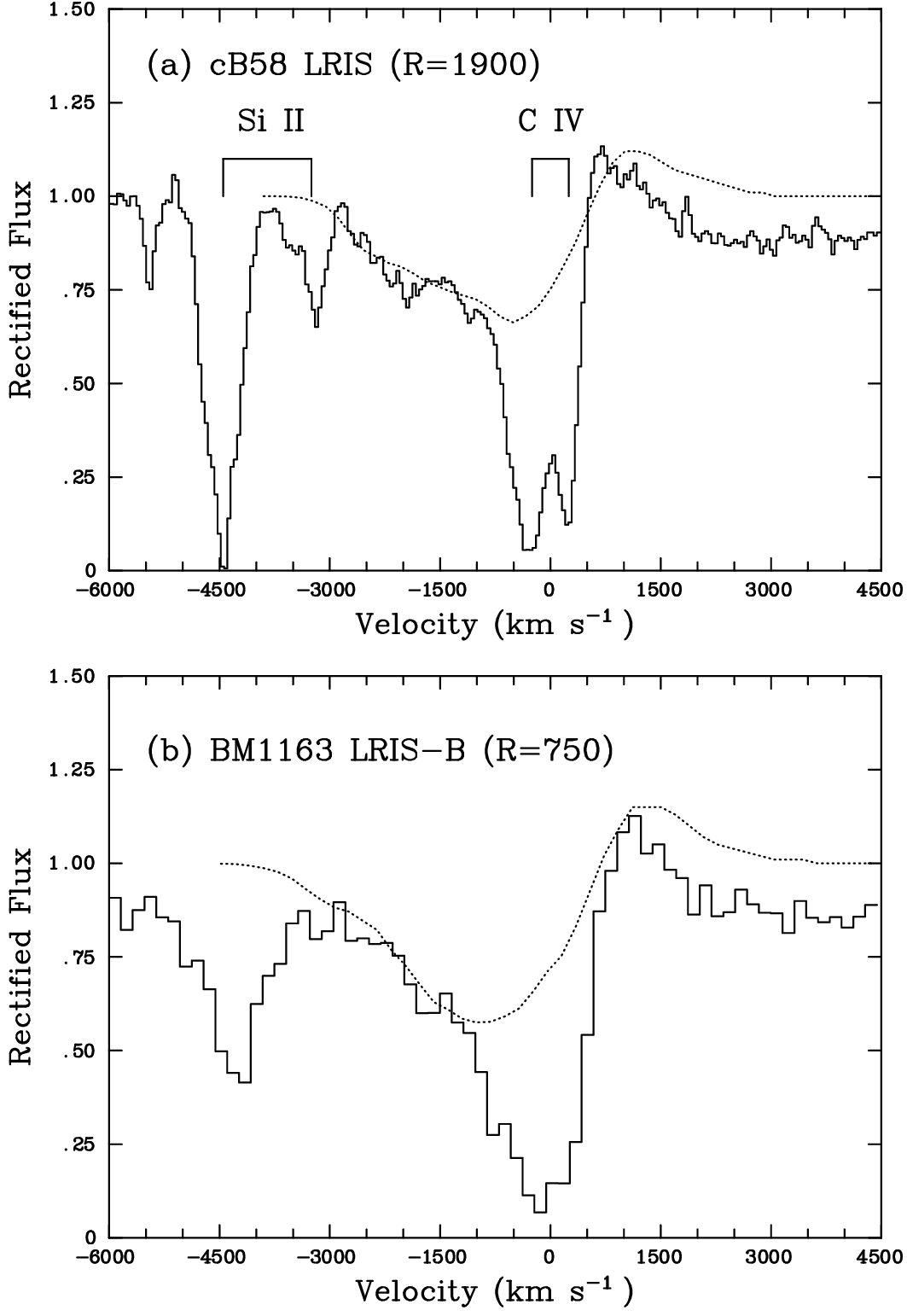}
\end{center}
\caption{Portions of the Keck I/LRIS spectrum ($R=1900$) of the 
$z = 2.7268$ gravitationally lensed galaxy MS\,1512$-$cB58 (top)
and of the Keck/LRIS-B spectrum ($R=750$) of the 
$z = 1.4105$ UV bright galaxy Q1307-BM1163 (bottom)
in the region encompassing the C\,{\sc iv}\,$\lambda 1549$ 
spectral feature. 
Each spectrum has been reduced to the rest frame of the galaxy
and normalised using the continuum windows identified by
Rix et al. (2004). In each panel we also show with a dotted line the 
Sobolev with Exact Integral fit to the stellar wind component of
C\,{\sc iv}, indicating terminal velocities
 $v_{\infty} = 3000$ and 3200\,km~s$^{-1}$
 in cB58 and BM1163 respectively. 
Narrower interstellar absorption lines of Si\,{\sc ii} and C\,{\sc iv}
are indicated with vertical tick marks. A blend of broad photospheric
absorption features depresses the continuum longward of the 
C\,{\sc iv} P Cygni emission peak. 
\label{medium}}
\end{figure}

Provided the stellar and interstellar components can be resolved,
it should still be possible to use the equivalent width of the former
to deduce the metallicity of the OB stars in which it arises,
if the local calibration derived here---and re-enforced
by the spectral modelling by Rix et al. (2004)---also applies to
high redshift star-forming galaxies.
We test this hypothesis by considering two high redshift
galaxies for which spectra of unusually high S/N are available
in the literature, the $z = 2.7268$ gravitationally lensed galaxy
MS\,1512$-$cB58 (Pettini et al. 2000, 2002) and the 
$z = 1.4105$ UV bright galaxy Q1307-BM1163 from the
survey of galaxies in the `redshift desert' by Steidel et al. (2004)
(see Fig.~\ref{medium}).

As can be readily appreciated from Fig.~\ref{medium},
the narrower interstellar lines make a major contribution
to the C\,{\sc iv} feature in both galaxies.
Consequently, we have used the Sobolev with Exact
Integral (SEI: Lamers, Cerruti-Sola, \& Perinotto 1987; 
Massa et al. 2003) technique 
to fit solely the wind component, shown by the dotted lines
in the figure. 
The SEI approach 
has been widely, and successfully, applied to synthesising
the spectral morphology of UV P Cygni stellar wind profiles
from OB stars, and as such is ideally suited to our purposes.
We measure EW(C\,{\sc iv})\,=\,3.8\,\AA\ and 5.5\,\AA\
for cB58 and BM1163, respectively. Using the calibration of
Eqn.~(1), we deduce corresponding oxygen abundances of
$\log {\rm (O/H)} + 12 = 8.17$ and 8.51 respectively.
These values in turn imply metallicities $Z \simeq 0.32$ and
$0.71 Z_{\odot}$ respectively, relative to the solar abundance 
$\log {\rm (O/H)} + 12 = 8.66$ (Asplund et al. 2004).

We now compare these values with those obtained from
other metallicity indicators available for these two galaxies.
We consider cB58 first. Nebular emission lines were measured
by Teplitz et al. (2000) who deduced 
$\log {\rm (O/H)} + 12 = 8.39$ using the $R23$ 
index of Pagel et al. (1979). From the emission
line fluxes listed in Table 1 of Teplitz et al.,  it can also be seen that
$N2 \equiv \log$\,([N\,{\sc ii}]\,$\lambda 6583/{\rm H}\alpha) = -1.04$
which implies $\log {\rm (O/H)} + 12 = 8.31$
using the calibration by Pettini \& Pagel (2004).
Furthermore, from 
$O3N2 \equiv \log$\{([O\,{\sc iii}]$\lambda 5007/{\rm H}\beta)$/([N\,{\sc ii}]\,$\lambda 6583/{\rm H}\alpha)$\}\,$ = 1.60$
we find $\log {\rm (O/H)} + 12 = 8.22$ from Eqn.~(3) of
Pettini \& Pagel (2004).
Each of these strong line abundance estimators has an accuracy 
of about $\pm 0.2$\,dex.

Using high resolution spectroscopy, Pettini et al. (2002) were able
to measure the abundances of several elements in the neutral
interstellar gas of cB58. They found that the alpha-capture elements
Mg, Si, P, and S have abundances of $\sim 0.4$ solar, while 
the Fe-peak elements Mn, Fe, and Ni are more underabundant,
at $\sim 0.1$ solar. In each case the uncertainty of these
determinations is about $\pm 0.1$\,dex. The difference
between the two groups of elements
may be real, reflecting the prompt release
of the nucleosynthetic products of massive stars, 
and/or it could be due to  dust depletion of the Fe-peak 
elements---probably both effects contribute.

Comparing all of these estimates with $\log {\rm (O/H)} + 12 = 8.17$
($Z \simeq 0.32 Z_{\odot}$) deduced here from EW(C\,{\sc iv}),
we conclude that the latter is in reasonable agreement with other
abundance indicators.
The same conclusion is reached for Q1307-BM1163 for which Steidel
et al. (2004) reported $\log {\rm (O/H)} + 12 = 8.53 \pm 0.25$
from the $N2$ index, in good agreement with 
$\log {\rm (O/H)} + 12 =  8.51$ we deduced from EW(C\,{\sc iv}), 
although our BM1163 stellar wind fit was in part guided by that
of cB58.

Clearly, in both cases discussed here we would have considerably 
\emph{over}estimated the metallicity 
had we not been able to resolve the wind component from the
interstellar absorption of C\,{\sc iv}$\lambda 1549$.
Specifically, the combined equivalent widths, measured 
within the narrow window on which Eqn.~(1) is based, are
7.0 and 8.8\,\AA, corresponding to 
$\log {\rm (O/H)} + 12 =  9.02$ and 9.50.
These are highly discrepant from all other metallicity measurements, 
as summarised in  Table~\ref{table}.

Before concluding this section, we point out that our SEI fits to the
stellar  P Cygni components of the C\,{\sc iv} lines in cB58 and BM1163 
are further supported by the results of spectral synthesis models.
Specifically, Pettini et al. (2003) presented Starburst99 (Leitherer
et al. 1999, 2001) fits to the spectral region near 1550\,\AA\ in these
galaxies obbtained using template spectra of OB stars of, respectively, 
Magellanic Cloud and Milky Way metallicity, and assuming continuous star
formation with a Salpeter IMF. The spectral morphology of our SEI fits
to the C\,{\sc iv}~$\lambda$1549 profiles in cB58 and BM1163 reproduced in
Fig.~\ref{medium} closely resembles the corresponding Starburst99 models.

%
%
\begin{figure}
\begin{center}
\includegraphics[width=0.9\columnwidth,clip,angle=0]{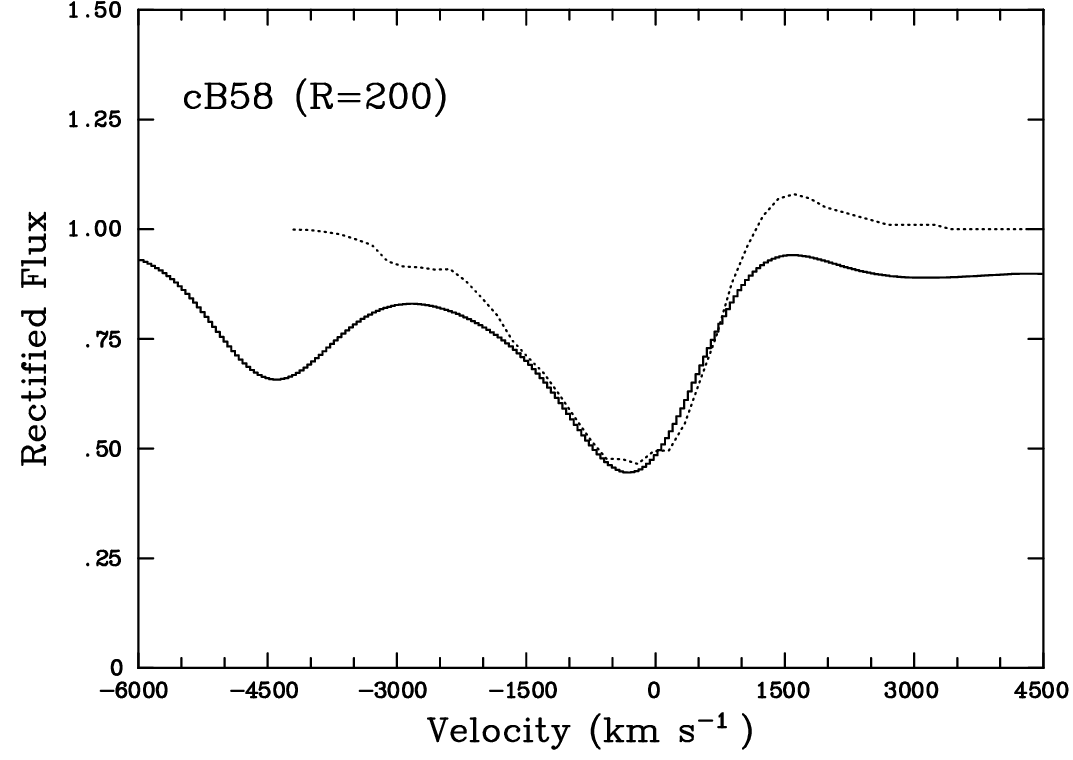} 
\end{center}
\caption{Portion of the Keck I/LRIS spectrum of cB58 degraded to a
resolving power $R$=200, typical of many surveys for high redshift
galaxies. At this coarse resolution even the Sobolev with Exact Integral
fitting technique (dotted line) is unable to separate stellar and
interstellar components of the C\,{\sc iv} line, leading to systematic
overestimates of the metallicity. \label{low}} 
\end{figure}

\subsection{Abundances from low resolution rest-frame UV spectroscopy}

The resolution of the LRIS-B spectrum of Q1307-BM1163 
shown in Fig.~\ref{medium}b ($R \simeq 750$)
is near the minimum required for 
deconvolving stellar and interstellar components of 
C\,{\sc iv}\,$\lambda 1549$. 
While this resolving power is typical of the surveys
for Lyman break, BX, and BM galaxies by Steidel et al. (2003, 2004),
many other surveys for high redshift galaxies have been conducted
at lower spectral resolutions. Thus, the FORS Deep Field survey,
on which the study of Mehlert et al. (2002) is based, employs
$R \simeq  200$; the Gemini Deep Deep Survey (GDDS) of 
Abraham et al. (2004) has $R \simeq 300$ in the blue (where
the C\,{\sc iv}\,$\lambda 1549$ line falls at the redshifts
of most GDDS galaxies); and the VIMOS VLT Deep Survey
(VVDS) of Le F\`{e}vre et al. 2005 delivers spectra with $R \sim 230$.

These resolutions are too coarse for the purpose of measuring
the equivalent width of the stellar C\,{\sc iv} line, as demonstrated 
in Fig.~\ref{low}. Once the LRIS spectrum of cB58 shown in 
Fig.~\ref{medium} is degraded to $R = 200$,
we measure EW(C\,{\sc iv})\,=5.8\,\AA\ by straightforward
integration across our 1534--1551\,\AA\ spectral window and
EW(C\,{\sc iv})\,=5.3\,\AA\ from the SEI fit.\footnote{Note 
that EW(C\,{\sc iv}) is not conserved, and is lower than the 
value we measured from the $R= 1900$ spectrum, 
due to the restricted integration
range of --2750 to +550 km\,s$^{-1}$ we use.}
Even adopting the
latter of the these two measurements we would still overestimate
the oxygen abundance at $\log {\rm (O/H)}+12 = 8.57$ from Eqn.~(1).
Evidently, such low resolution data are inadequate for measuring
abundances via the C\,{\sc iv}$\lambda 1549$ line (see Table~\ref{table}). 

%
%

\begin{table}
\caption{Summary of various metallicity ($\log $(O/H)+12)
indicators for cB58 and BM1163 considered here, 
including strong nebular 
line methods ($R23$, $N2$ and $O3N2$, see text) 
and application of Eqn. (1)
based on the measured EW(C\,{\sc iv}) either excluding (SEI 
model) or including interstellar contributions.}\label{table}
\begin{tabular}{l@{\hspace{0.25mm}}r@{\hspace{2mm}}
c@{\hspace{1.75mm}}c@{\hspace{1mm}}c@{\hspace{1mm}}c@{\hspace{1mm}}c}
\hline
Galaxy              &$R~$          & $R23$         & $N2$        & $O3N2$  
& ~EW(C\,{\sc iv}) & EW(C\,{\sc iv}) \\ 
                         &                  &                   &                 &                
& Wind                 & Wind+ISM \\
\hline
MS1512--cB58    & 1900          & 8.39           & 8.31    & 8.22       & 
8.17   &  9.02  \\
                          &  200           &                  &            &   
           & 8.67   &  8.79 \\ 
Q1307--BM1163  & 750            & ---             & 8.53      &    ---       
& 8.51    &  9.50  \\
\hline
\end{tabular}
\end{table}

\section{Summary and Conclusions}

The C\,{\sc iv}\,$\lambda 1549$ line, one of the strongest
features in the rest-frame UV spectra of star-forming
galaxies, may well be the only tool at our disposal for
measuring the degree of chemical evolution of galaxies
at $z \ga 3.5$, at least until the advent of ground-based
30\,m telescopes and of space-bourne near-IR spectrographs
on large aperture telescopes such as the \emph{JWST}.
This realisation has prompted us to reconsider the usefulness
and limitations of this transition as an abundance indicator.
Our main conclusions are as follows.

1.  C\,{\sc iv}\,$\lambda 1549$  is a complex spectral feature
which in starburst galaxies consists of a blend of stellar
P Cygni emission-absorption, narrower interstellar absorption,
and potentially nebular emission, although this last component
is of minor importance except in galaxies with Active Galactic Nuclei (AGN)
(e.g. Leitherer, Calzetti, \& Martins 2002). 
The same considerations apply to Si\,{\sc iv}\,$\lambda 1397$
which is however normally weaker, and consequently less
accurately measured, than C\,{\sc iv}\,$\lambda 1549$.
Thus, we see little practical advantage in averaging the equivalent
widths of the two lines to give a combined
measure of the strength of wind absorption,
as done by Heckman et al. (1998), and instead consider it
advantageous to concentrate on  
C\,{\sc iv}\,$\lambda 1549$ alone, as proposed by Mehlert et al. (2002).

2. In high redshift star-forming galaxies, the interstellar component
of C\,{\sc iv} can be the dominant contributor to the composite
spectral feature. It is therefore essential, for 
abundance determinations, to obtain spectra of sufficiently
high resolution for the stellar and interstellar components
to be clearly recognised---$R \simeq 1000$ seems to be the minimum
resolving power required.
A fitting technique such as the Sobolev with Exact Integral
method of Lamers et al. (1987) accounts for the stellar 
wind profile morphology.
Without it, the strength of the wind line---and the metal abundance
it implies---would be systematically overestimated.
Furthermore, we advocate measuring the equivalent width of 
C\,{\sc iv}\,$\lambda 1549$, EW(C\,{\sc iv}), over a restricted
wavelength range, from 1534 to 1551\,\AA, corresponding
to wind velocities in the interval $-2750$ to +550\,km~s$^{-1}$,
thereby avoiding contamination with 
Si\,{\sc ii}\,$\lambda\lambda 1527, 1533$ interstellar absorption.

3. We have derived a local calibration between EW(C\,{\sc iv}), 
measured between 1534 and 1551\,\AA, and the oxygen
abundance (from the $R23$ method):
\[ \log {\rm (O/H)} + 12  = 7.15 + {\rm EW(C\,IV)}/3.74 \]
based on low resolution \emph{IUE} spectra of nearby starburst galaxies.
While the \emph{IUE} spectra do not resolve stellar and interstellar
components, the above relationship is in very good agreement
with the fully synthetic starburst spectra computed by
Rix et al. (2004). Given that the latter are purely stellar, 
it would appear that in local star-forming galaxies---unlike 
their high $z$ counterparts---interstellar C\,{\sc iv} absorption 
makes a relatively minor contribution to the blend.

4. We applied the above calibration to the stellar 
component of the C\,{\sc iv} line in two well-observed
high redshift star-forming galaxies---the gravitationally
lensed $z = 2.7268$  galaxy MS\,1512$-$cB58 and the 
$z = 1.4105$ UV bright galaxy Q1307-BM1163.
The resulting abundances, $12 + \log ({\rm O/H)} = 8.2$ and 
8.5 respectively, are in encouraging agreement with
those determined by other methods, within the 
$\sim$0.2\,dex uncertainties of other abundance determinations.

5. Even so, the local relationship between EW(C\,{\sc iv})
and the oxygen abundance shown in Fig.~\ref{mean}b
exhibits a worryingly large scatter which demands 
further study. 

One problem is the determination
of a trustworthy oxygen abundance in the galaxies:
values of $12 + \log({\rm O/H)} > 9.0$ 
($Z \ga 2 Z_{\odot}$) in Fig.~\ref{mean}b are called
into question by recent work which has failed to confirm
supersolar abundances in individual H\,{\sc ii} regions
of nearby spirals (Bresolin et al. 2005 and references
therein). If the $R23$ index systematically
overestimates (O/H) at high metallicities, the 
relationship shown in Fig.~\ref{mean}b may
be altered significantly, rather than requiring a simple
systematic downward revision.

A second cause for concern is the fact that EW(C\,{\sc iv})
responds not only to metallicity, but also to the age of the
starburst, as vividly demonstrated by the difference
between `field' and `super star cluster' UV light in
nearby galaxies (Chandar et al. 2005). However, this difficulty
may be alleviated when considering the integrated 
spectrum of a whole galaxy, for which the idealised case
of continuous star formation is more likely to be an
adequate approximation to reality.

Thus, in order to improve on the present calibration of 
EW(C\,{\sc iv}) vs. (O/H) we require observations
of starburst galaxies with sufficiently high \emph{spectral} 
resolution to separate stellar and interstellar lines, 
but of sufficiently coarse \emph{spatial} resolution to give
an integrated spectrum truly representative of the whole galaxy.
Furthermore, such observations should be performed in galaxies 
where a variety of abundance indicators are available, including
at least $N2$ as well as $R23$.
Paradoxically, these seemingly mutually exclusive requirements
can at present be met most easily in galaxies at 
$z \simeq 1.5 - 2.5$ observed from the ground with optical
and near-IR spectrographs on large telescopes.
Certainly, a relationship between 
EW(C\,{\sc iv}) and (O/H) determined for galaxies at 
$z \simeq 1.5 - 2.5$  would be the appropriate baseline
against which to consider galaxies at higher redshifts.
Once \emph{HST} is equipped again 
with a working UV spectrograph,
it may become possible to examine such a relationship
for much nearer galaxies drawn from large data bases,
as produced for example by the Sloan Digital Sky Survey
(e.g. Tremonti et al. 2004; Gallazzi et al. 2005).

\section*{Acknowledgments}

Based, in part, on observations with the NASA/ESA Hubble Space Telescope, 
obtained from the ESO/ST-ECF Science Archive Facility. PAC 
gratefully acknowledges financial support from the Royal Society. 
We thank Sam Rix for providing her synthetic UV spectra of star-forming
galaxies.


\begin{thebibliography}{}

\bibitem[]{} Abraham, R.~G., Glazebrook, K., McCarthy, P.~J., et al.,  
2004, AJ, 127, 2455

\bibitem[]{} Asplund M., Grevesse N.,, Sauval A.~J., Allende Prieto C., 
Kiselman D. 2004, A\&A, 417, 751

\bibitem[]{} Bresolin F., Garnett D.~R., Kennicutt R.~C., 2004, ApJ, 615, 228

\bibitem[]{} Bresolin F., Schaerer, D., Gonz‡lez Delgado, R.~M., 
Stasinska, G., 2005, A\&A, 441, 981

\bibitem[]{} Bunker, A.~J., Stanway, E.~R., Ellis, R.~S., McMahon, R.~G. 2004,
MNRAS, 355, 374

\bibitem[]{} Chandar R., Leitherer C., Tremonti C.~A., 2004, ApJ, 604, 153

\bibitem[]{} Chandar, R., Leitherer, C., Tremonti, C.~A., 
Calzetti, D., Aloisi, A., Meurer, G.~R., de Mello, D., 2005, ApJ, 628, 210 


\bibitem[]{} Erb D.~K., Shapley A.~E., Pettini M., Steidel C.~C., Reddy 
 N.~A., Adelberger K.~L., 2006, ApJ, submitted

\bibitem[]{} Gallazzi, A., Charlot, S., Brinchmann, J., White, S.~D.~M.,
Tremonti, C.~A., 2005, MNRAS, 362, 41

\bibitem[]{} Garnett, D.~R., Edmunds, M.~G., Henry, R.~B.~C., Pagel, B.~E.~J., 
Skillman, E.~D., 2004a, AJ, 128, 2772

\bibitem[]{} Garnett, D.~R., Kennicutt, R.~C., Bresolin, F., 2004, ApJ, 607, L21

\bibitem[]{} Heckman T.~M., Robert C., Leitherer C., Garnett D.~R., 
van der Rydt F., 1998, ApJ, 503, 646

\bibitem[]{} Hopkins,  A.~M., 2004, ApJ, 615, 209

\bibitem[]{} Howarth I.~D., Murray J., Mills D., Berry D.~S., 2003, 
Starlink User Note 50.24, Rutherford Appleton Laboratory


\bibitem[]{} Iwata, I., Ohta, K., 
Tamura, N., Ando, M., Akiyama, M.,  Aoki, K., 2005, 
in de Grijs, R., Gonzalez Delgado, R.~M., eds,
Starbursts: From 30 Doradus to Lyman Break Galaxies.
Springer-Verlag, Berlin, p.~29

\bibitem[]{} Kewley, L.~J., Dopita, M.~A., 2002, ApJS, 142, 35 
 
\bibitem{} Kewley, L., \& Kobulnicky, H.~A. 2005, 
in de Grijs, R., Gonzalez Delgado, R.~M., eds,
Starbursts: From 30 Doradus to Lyman Break Galaxies.
Springer-Verlag, Berlin, p.~307

\bibitem[]{} Kinney A., Bohlin R., Calzetti D., Panagia N., Wyse R.~F.~G., 1993, 
ApJS, 86, 5

\bibitem[]{} Lamers H.~J.~G.~L.~M., Cerruti-Sola M., Perinotto M., 1987, ApJ, 314, 726

\bibitem[]{} Leitherer C., Calzetti D., Martins L.~P., 2002, ApJ, 574, 114 

\bibitem[]{} Leitherer C., Schaerer D., Goldader J.D. et al. 1999, ApJS 
123, 3

\bibitem[]{} Leitherer C., Le\~ao J.~R.~S., Heckman T.~M., Lennon D.~J., Pettini 
M., Robert C., 2001, ApJ, 550, 724

\bibitem[]{} Le F\`{e}vre, O., Vettolani, G., Garilli, B., et al., 2005, Nature, 437, 519

\bibitem[]{} Madau P., Ferguson H.~C., Dickinson M. et al., 1996, MNRAS, 283, 1388

\bibitem[]{} Massa D., Fullerton A.~W., Sonneborn G., Hutchings J.~B., 2003, 
ApJ, 586, 996

\bibitem[]{} Mehlert D., Noll S., Appenzeller I. et al., 2002, A\&A, 393, 809

\bibitem[]{} Mehlert D., Tapken C., Appenzeller I., Noll S., de Mello, D.,
Heckman T.~M., 2005, in de Grijs, R., Gonzalez Delgado, R.~M., eds,
Starbursts: From 30 Doradus to Lyman Break Galaxies.
Springer-Verlag, Berlin,  p.~299

\bibitem[]{} Pagel B.~E.~J., Edmunds M.~G., Blackwell D.~E., Chun M.~S., Smith 
M.~G., 1979, MNRAS, 189, 95

\bibitem[]{} Pettini M., 2006, in Le Brun, V., Mazure, A., Arnouts, S., Burgarella, D.,
eds, The Fabulous Destiny of Galaxies: Bridging Past and Present.
Edition Fronti\`{e}res, Paris, in press

\bibitem[]{} Pettini M., Pagel, B.~E.~J. 2004, MNRAS, 348, L59 

\bibitem[]{} Pettini M., Steidel C.~C., Adelberger K.~L., Dickinson M., 
Giavalisco M., 2000, ApJ, 528, 96

\bibitem[]{} Pettini M., Rix S.~A., Steidel C.~C., Adelberger K.~L., Hunt 
M.~P., Shapley A.~E., 2002, ApJ, 569, 742

\bibitem[]{} Pettini M., Rix S.~A., Steidel C.~C., Shapley A.~E., 
Adelberger K.~L., 2003, in van der Hucht K.~A., Herrero A., Esteban, 
C., eds, A Massive Star Odyssey, from Main Sequence to Supernova, Proc 
IAU Symp 212. ASP, San Francisco, p.~671

\bibitem[]{} Prinja R.~K., Crowther P.~A., 1998, MNRAS, 300, 828

\bibitem[]{} Rix S.~A., Pettini M., Leitherer C., Bresolin F., Kudritzki 
R-P., Steidel C.~C., 2004, ApJ, 615, 98


\bibitem[]{} Shapley A.~E., Steidel, C.~C., Pettini, M.,  Adelberger, K.~L., 2003,
ApJ, 588, 65


\bibitem[]{} Steidel, C.~C., 
Adelberger, K.~L., Shapley, A.~E., Pettini, M., Dickinson, M.,  
Giavalisco, M., 2003, ApJ, 592, 728 

\bibitem[]{} Steidel C.~C., Shapley  A.~E., Pettini M., Adelberger K.~L., Erb D.~K.,
Reddy N.~A., Hunt M.~P., 2004, ApJ, 604, 534

\bibitem[]{} Storchi-Bergmann, T., Calzetti, D., Kinney, A., 1994, ApJ, 429, 572

\bibitem[]{} Teplitz H.~I., McLean I.~S., Becklin E.~E. et al,  2000, ApJ, 533, L65

\bibitem[]{} Tremonti, C.~A., Heckman, T.~M., Kauffmann, G., et  al., 2004, 
ApJ, 613, 898 

\bibitem[]{} Vink J.~S., de Koter A., Lamers H.~J.~G.~L.~M., 2001, A\&A, 369, 574

\end{thebibliography}
\end{document}